\documentclass[%
 aip,
 amsmath,amssymb,
 reprint,%
]{revtex4-1}

\usepackage{graphicx}
\usepackage{dcolumn}
\usepackage{bm}

\usepackage[utf8]{inputenc}
\usepackage[T1]{fontenc}
\usepackage{mathptmx}
\usepackage{etoolbox}

\usepackage{siunitx} 
\usepackage[version=3]{mhchem}
\usepackage{chemformula}

\newcommand{\br}{\bm{r}}

\newcommand{\s}{_\mathrm{{\scriptscriptstyle S}}}
\newcommand{\h}{_\mathrm{{\scriptscriptstyle H}}}
\newcommand{\xc}{_\mathrm{{\scriptscriptstyle XC}}}

\newcommand{\ext}{_\mathrm{{\scriptscriptstyle ext}}}

\newcommand{\fwu}{%
    Center for Advanced Systems Understanding, 
    D-02826 G\"orlitz, 
    Germany}
\newcommand{\hzdr}{%
    Helmholtz-Zentrum Dresden-Rossendorf, 
    D-01328 Dresden, 
    Germany}
\newcommand{\snl}{%
    Sandia National Laboratories, 
    Albuquerque, NM 87185, 
    USA}

\makeatletter
\def\@email#1#2{%
 \endgroup
 \patchcmd{\titleblock@produce}
  {\frontmatter@RRAPformat}
  {\frontmatter@RRAPformat{\produce@RRAP{*#1\href{mailto:#2}{#2}}}\frontmatter@RRAPformat}
  {}{}
}%
\makeatother
\begin{document}

\preprint{AIP/123-QED}

\title{Impact of Electronic Correlations on High-Pressure Iron: Insights from Time-Dependent Density Functional Theory}
\author{K. Ramakrishna}
\altaffiliation{Electronic mail: k.ramakrishna@hzdr.de}
\affiliation{\fwu}
\affiliation{\hzdr}%

\author{M. Lokamani}%
\affiliation{\hzdr}%

\author{A. Baczewski}
\affiliation{\snl}%

\author{J. Vorberger}%
\affiliation{\hzdr}%

\author{A. Cangi}%
\altaffiliation{Electronic mail: a.cangi@hzdr.de}
\affiliation{\fwu}
\affiliation{\hzdr}%

\date{\today}

\begin{abstract}
We present a comprehensive investigation of the electrical and thermal conductivity of iron under high pressures at ambient temperature, employing the real-time formulation of time-dependent density functional theory (RT-TDDFT). Specifically, we examine the influence of a Hubbard correction (+\textit{U}) to account for strong electron correlations. Our calculations based on RT-TDDFT demonstrate that the evaluated electrical conductivity for both high-pressure body-centered cubic (BCC) and hexagonal close-packed (HCP) iron phases agrees well with experimental data. Furthermore, we explore the anisotropy in the thermal conductivity of HCP iron under high pressure, and our findings are consistent with experimental observations. Interestingly, we find that the incorporation of the +\textit{U} correction significantly impacts the ground state and linear response properties of iron at pressures below 50 GPa, with its influence diminishing as pressure increases. This study offers valuable insights into the influence of electronic correlations on the electronic transport properties of iron under extreme conditions.  
\end{abstract}

\maketitle

\section{\label{sec:level1}Introduction}

The properties of elemental iron under high pressure are of great interest due to its prominence in Earth's core~\cite{anderson1986properties,doi:10.1063/1.1708965,anzellini2013melting,kraus2022measuring}. Two solid phases of iron have been broadly studied. Under ambient conditions, the stable structure of iron is a single-atom body-centered cubic (BCC) cell. At room temperature and pressures relevant to Earth's core (360~GPa)~\cite{kuwayama2008phase,tateno2010structure}, a single-atom hexagonal close-packed (HCP) cell appears to be the stable structure~\cite{PhysRevLett.97.215504, tateno2010structure, mao1990static}. The phase boundary between the two phases depends on the magnetic ordering in the BCC phase. Ferromagnetic BCC is the energetically favored ground state compared to paramagnetic BCC ~\cite{zhang2011density}. Electronic correlations are significant in iron but become less significant under pressure because the electronic kinetic energy increases more rapidly than the potential energy.

Experiments using x-ray magnetic circular dichroism and x-ray absorption spectroscopy indicate a BCC-HCP transition~\cite{bancroft1956polymorphism,takahashi1964high} between 12-18~GPa. The disappearance of the net magnetic moment indicates the HCP phase is non-magnetic~\cite{PhysRevLett.93.255503}. As the pressure increases, 3$d$ electrons delocalize further and form hybridized 3$d$-4$p$ states. This goes along with a reduction in magnetic moment~\cite{iota2007electronic}. 

Material properties are commonly modeled using density functional theory (DFT)\cite{PhysRev.136.B864, PhysRev.140.A1133}. While successful, strict semi-local approximations typically fail to capture even the qualitative physics of strongly correlated materials. However, methods that incorporate a Hubbard correction (DFT+$U$) for the on-site Coulomb interaction have found great success in computing the structural and electronic properties of these challenging materials ~\cite{anisimov1997first,cococcioni2005linear}. Dynamical mean field theory (DMFT)+DFT is an alternative approach for augmenting conventional DFT, which treats electronic correlations differently~\cite{RevModPhys.78.865,paul2019applications,sanchez2009strength,sponza2017self}.

The standard approach for computing the electrical conductivity uses the Kubo-Greenwood (KG) formula~\cite{doi:10.1143/JPSJ.12.570,Greenwood_1958}. Commonly, it is evaluated on Kohn-Sham (KS) orbitals, eigenvalues, and occupation numbers. This method has been widely used to compute the electrical conductivity in various systems, including strongly coupled ~\cite{PhysRevE.66.025401,PhysRevLett.118.225001,PhysRevB.106.054304}, as well as liquid and solid metals~\cite{PhysRevB.84.054203,de2012electrical,PhysRevB.85.184201,PhysRevLett.122.086601,PhysRevB.90.165113,PhysRevB.107.085145}. 
However, employing KS quantities in the KG formula neglects the interaction kernel, which is crucial for capturing collective effects that govern electronic excitations and determine transport properties~\cite{pourovskii2017electron}.

A more accurate alternative that captures electronic excitations is time-dependent density functional theory (TDDFT). Within TDDFT, the standard approach is to employ linear-response TDDFT (LR-TDDFT), which calculates the interacting response function using an interaction kernel that incorporates electron correlations through the Hartree and exchange-correlation (XC) kernel. The effectiveness of LR-TDDFT in computing transport properties with full wavenumber and frequency resolution has been extensively studied for both solid materials~\cite{PhysRevB.84.075109} and liquid aluminum~\cite{ramakrishna2020firstprinciples}.
An alternative method is real-time propagation in time-dependent density functional theory (RT-TDDFT)\cite{PhysRevB.54.4484,baczewski2016x,baczewski2021predictions}. RT-TDDFT directly simulates the microscopic form of Ohm's law by obtaining the time-dependent physical current from the system's linear response to an electric field. According to Ohm's law, the electrical conductivity can be directly computed from the induced time-dependent current~\cite{PhysRevB.107.115131,10.1063/5.0138955}.

While TDDFT captures the transport properties of simple metals like aluminum quite well, its performance for less simple metals has not yet been established. DFT+$U$ has been traditionally used in evaluating static properties~\cite{cococcioni2005linear}, but the Hubbard correction has not yet been included in the TDDFT calculations of dynamical properties such as the electrical conductivity. 

In this work, we investigate the influence of electronic correlations on iron under pressure. Initially, we calculate the equation of state (EOS) using DFT and incorporate Hubbard corrections. By investigating the static properties, we establish a foundation for computing electronic transport properties. To evaluate the electrical conductivity of BCC and HCP iron under high pressure, we employ RT-TDDFT. We specifically account for strong electronic correlations by incorporating a Hubbard correction, allowing us to assess its impact on the computed conductivity.
Additionally, utilizing this methodology, we investigate the thermal conductivity and its anisotropy of HCP iron under high pressure, covering pressure conditions relevant to the Earth's core-mantle boundary (up to 150 GPa). Moreover, we demonstrate agreement of our RT-TDDFT results with the experimental anisotropy measurements reported by Ohta \textit{et al.}~\cite{ohta2018experimental}.

\section{Methodological and Computational Details}
Ground-state DFT calculations of the EOS and generation of the initial KS orbitals are performed using the full-potential linearized augmented plane-wave code implemented in Elk~\cite{elk}. A \emph{k}-point grid of $20\times 20\times 20$ points and $16\times 16\times 16$ points up to $60$ bands per atom are considered for the BCC and HCP phases, respectively in a unit cell. The $c/a$ ratio is fixed at 1.60 throughout for HCP~\cite{vekilova2015electronic,hausoel2017local} in the experimental range~\cite{PhysRevLett.110.117206,PhysRevLett.97.215504} as the variation with pressure is weak at higher pressures~\cite{steinle2004magnetism,steinle2004magnetism_jpcm}. The Perdew-Burke-Ernzerhof (PBE)~\cite{perdew1996generalized} exchange-correlation (XC) functional is used. The calculations with the Hubbard correction are also carried out with spin-polarized PBE around the mean-field (AMF)~\cite{petukhov2003correlated} form with double counting~\cite{PhysRevB.49.14211,cococcioni2005linear} with an enhanced description of magnetic and structural properties (see Table. 1 in Appendix \ref{app}). The Hubbard parameters ($U=0.125$~Ha and $J=0.033$~Ha) used in this work are well-tested for high-pressure iron~\cite{PhysRevLett.110.117206,mohammed2010stability,vekilova2015electronic} and in the range used in earlier studies of ambient iron~\cite{cococcioni2005linear}.

We utilize RT-TDDFT~\cite{PhysRevB.54.4484} to compute the electrical conductivity based on the microscopic form of Ohm's law 
\begin{equation}
j_\alpha(\br, t) = \sum_\beta \int d\br' \int_{-\infty}^{t} dt'\ \sigma_{\alpha\beta}(\br,\br',t-t')\, E_\beta(\br',t')\,,
\label{eq.ohmslaw}
\end{equation}
where an external electric field $\boldsymbol{E}(\br', t) = -\partial \boldsymbol{A}(\br', t)/\partial t$ induces a current density $\boldsymbol{j}(\br, t)$, with $\boldsymbol{A}(\br', t)$ denoting the associated external vector potential.
To this end, we solve the time-dependent KS equations 
\begin{equation}
\label{eq:tdks}
\hat{H}\s \psi_{n,k}(\br,t)=i \frac{\partial}{\partial t} \psi_{n,k}(\br,t)
\end{equation}
for the KS orbitals $\psi_{n,k}(\br,t)$ with the effective Hamiltonian
\begin{equation}
\hat{H}\s = \frac{1}{2} \left[-i \nabla + \boldsymbol{A}\s(\br,t)\right]^{2} + v\s(\br,t)\,,
\end{equation}
where $v\s(\br,t)=v\ext(\br,t)+v\h(\br,t)+v\xc(\br,t)$ is the KS potential which is a sum of the external, the Hartree, and XC potentials, while the effective vector potential $\boldsymbol{A}\s(\br,t)=\boldsymbol{A}(\br,t) + \boldsymbol{A}\xc(\br,t)$ comprises the sum of the external vector potential and the XC contribution~\cite{vignale1996current,vignale1997time}. Also, note that we use Hartree atomic units throughout.
Solving the time-dependent KS equations yields the induced current density~\cite{vignale1996current} as
\begin{equation}
\boldsymbol{j}(\br,t)=\Im \left[\sum_{i}^{N} \psi_{n,k}^{*}(\br,t) \nabla \psi_{n,k}(\br,t)\right] + n(\br,t) \boldsymbol{A}\s(\br,t)\ .
\end{equation}

Assuming the electric field is constant in space and its time-dependence is treated within the dipole approximation, we obtain the dynamical electrical conductivity as 
\begin{equation}
\sigma_{\alpha\beta}(\omega) = \frac{\tilde{j}_\alpha(\omega)}{\tilde{E}_{\beta}(\omega)} \,,
\label{eq.ohmslaw.ft}
\end{equation}
where we integrated Eq.~(\ref{eq.ohmslaw}) over the spatial coordinate and performed Fourier transforms over the time variable with 
$\tilde{j}_{\alpha}(\omega) = \int dt \int d\br\  j_\alpha(\br,t) e^{i \omega t}$ and $\tilde{E}_{\beta}(\omega) = \int dt\  E_\beta(t) e^{i \omega t}$.
Note that both the current and electric field are vectors, while the electrical conductivity is a tensor.
To achieve a sufficient resolution in frequency space and to guarantee that the electric current eventually reaches a stable state, the duration of the real-time propagation is crucial. The energy window of the simulation is inversely related to the full-width half maximum (FWHM) of the pulse, and the spectral energy resolution is inversely proportional to the total propagation time. 

\begin{figure*}[t]  
\centering       
\includegraphics[width=0.85\linewidth]{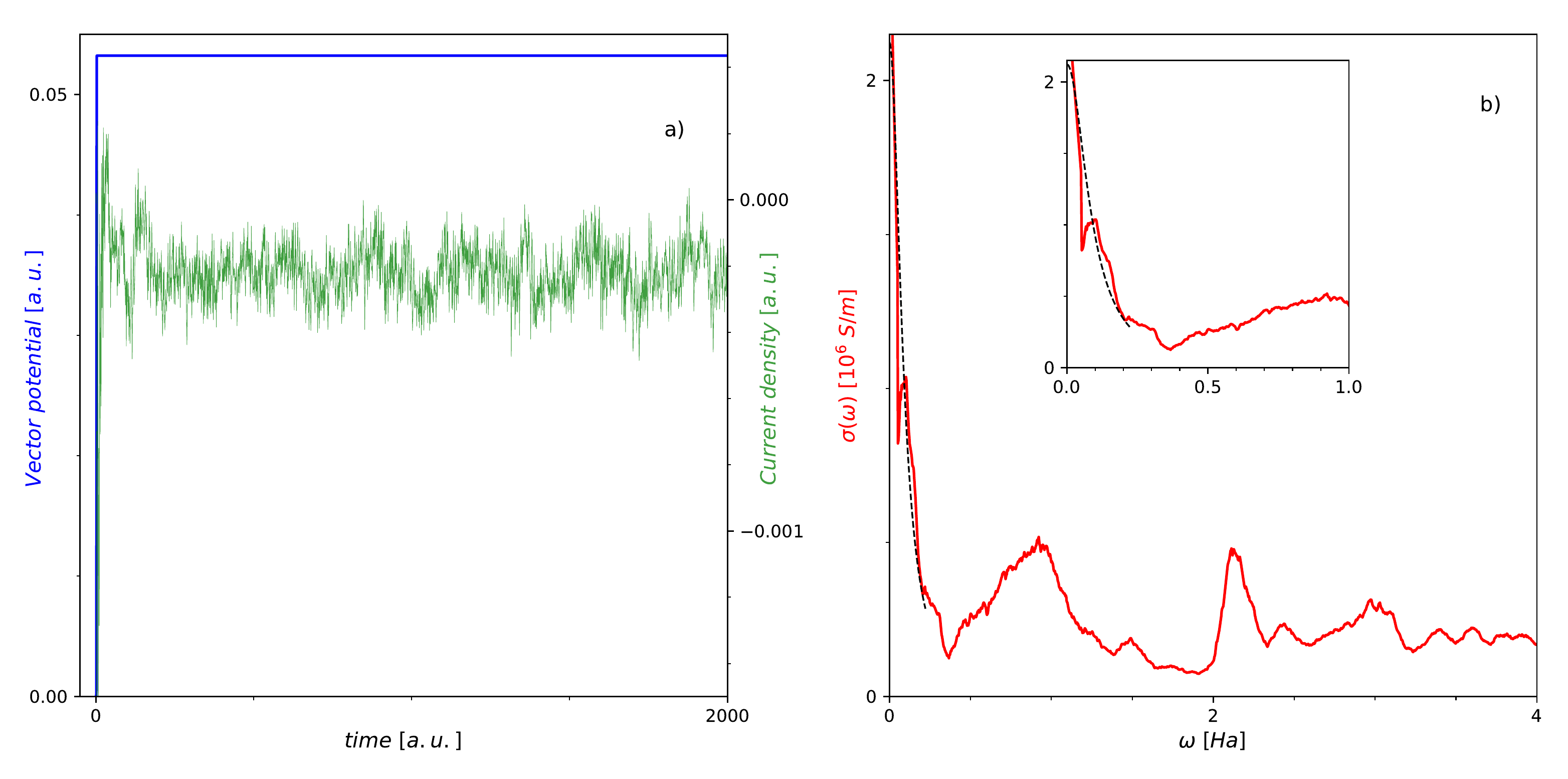}  
\caption{\raggedright a) Sigmoidal pulse (blue) in the $z$ direction vs. time with the induced current density (green) along $z$ vs. time, b) dynamical electrical conductivity component $\sigma_{zz}$. The black dashed line is the Drude fit that yields the DC conductivity. The inset plot shows the dynamical electrical conductivity at low energies.}
\label{rt_theory}  
\end{figure*}

Within the RT-TDDFT calculations, a sigmoidal pulse of vector amplitude 0.1~a.u. is applied with a peak time of 2~a.u. having an FWHM of 0.5~a.u. for a total simulation time of up to 2000~a.u (1 a.u.=24.819 attoseconds) providing an energy resolution up to 0.5~\si{\milli Ha} in frequency space. The pulse mentioned above (blue solid) is applied in the $z$ direction shown in Fig. \ref{rt_theory}a) with the induced current density (green solid) along $z$. The dynamical electrical conductivity (red solid) obtained from Eq.~(\ref{eq.ohmslaw.ft}) is shown in Fig.~\ref{rt_theory}b) where the DC conductivity is obtained from a Drude fit (black dashed) at low energies.

\section{Results}

\subsection{Equation of State}
\begin{figure*}   
\centering       
\includegraphics[width=0.95\linewidth]{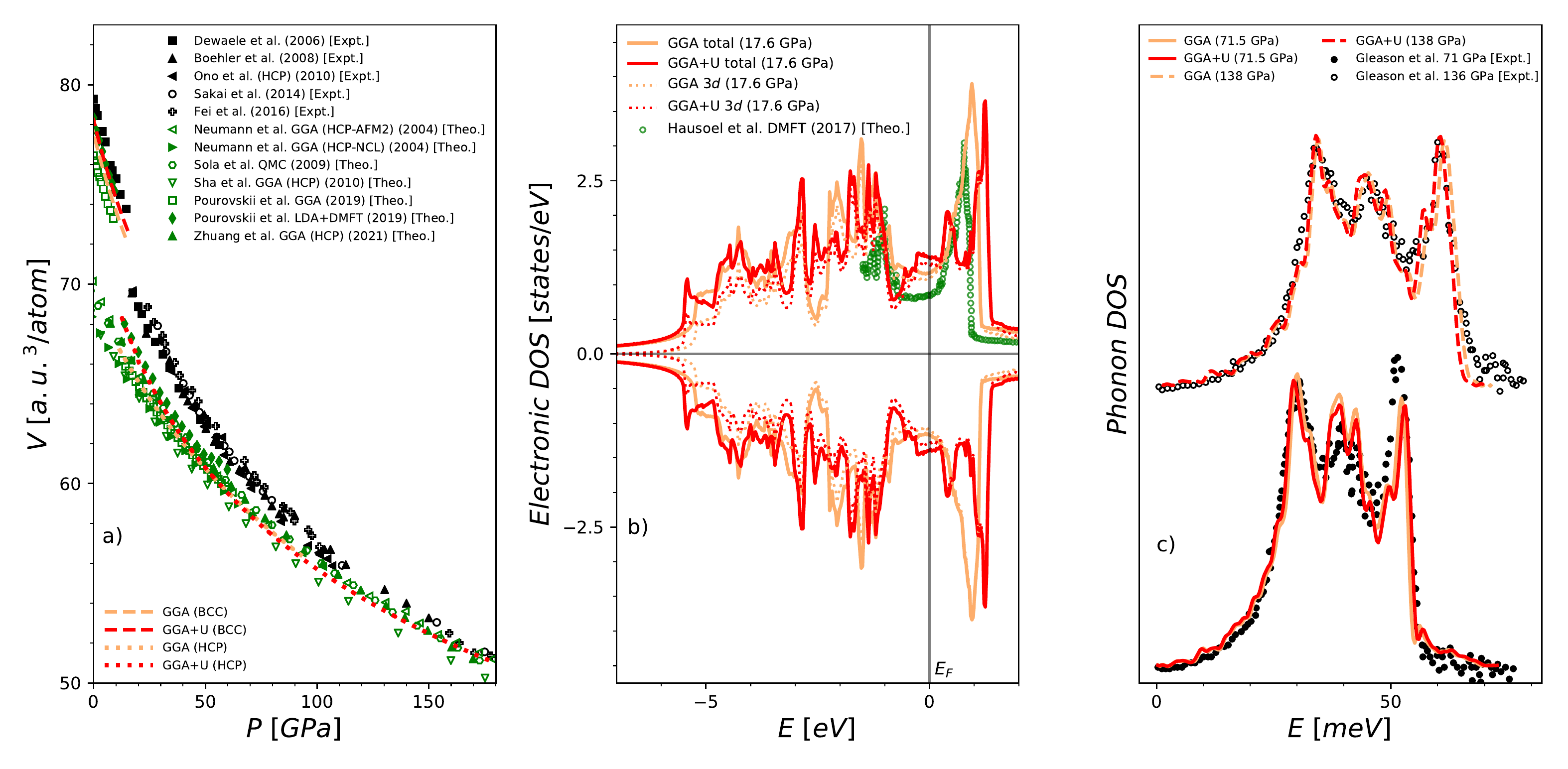}
\caption{a) Equation of state (EOS) for BCC and HCP iron. Experimental data for ambient temperature (T=300~K) stems from Refs.~\onlinecite{boehler2008melting,PhysRevLett.97.215504,sakai2014equation,fei2016thermal,ono2010high}. Theoretical data at T=0~K and ambient temperature (T=300~K) stems from Refs.~\onlinecite{pourovskii2019electronic,PhysRevB.81.094105,zhuang2021thermodynamic,sola2009equation,steinle2004magnetism_jpcm}, b) Spin-resolved density of states (DOS) of HCP iron at 17.6~GPa computed using DFT and DFT+$U$. Contribution from the 3\textit{d} states is indicated by the dotted lines. HCP DOS at ambient pressure stems from Ref.~\onlinecite{hausoel2017local}, c) Phonon density of states for HCP iron at 71.5~GPa (bottom) and 138~GPa (top) computed using DFT and DFT+$U$. Experimental data on HCP iron for pressures 71~GPa and 136~GPa at ambient temperature stems from Ref.~\onlinecite{gleason2013sound}. 
}     
\label{eos}
\end{figure*}

By employing ground-state DFT in conjunction with the Hubbard correction, we examine, as shown in Fig.~\ref{eos}, the EOS, the electronic density of states (DOS), and the phonon DOS of iron across a pressure range of approximately 180 GPa. In doing so, we specifically analyze the impact of electronic correlations on the EOS. In Fig.~\ref{eos}a), we present our findings (depicted as red and orange curves) alongside previous experimental measurements~\cite{boehler2008melting,PhysRevLett.97.215504,sakai2014equation,fei2016thermal,ono2010high} (denoted by black symbols) and theoretical predictions~\cite{pourovskii2019electronic,PhysRevB.81.094105,zhuang2021thermodynamic,sola2009equation,steinle2004magnetism_jpcm} (denoted by green symbols).
Our results are obtained by fitting total energies from DFT calculations using the PBE functional (GGA)~\cite{perdew1996generalized} for various volumes to the Vinet EOS~\cite{vinet1987temperature, vinet1989universal}. The calculations with a Hubbard correction (+$U$) are also carried out for the spin-polarized case using 
the PBE functional around the mean-field form with double counting~\cite{PhysRevB.49.14211,cococcioni2005linear}. 

We first examine the equation of state (EOS) shown in Fig. \ref{eos}a). In the BCC phase, the inclusion of the GGA$+U$ correction yields a relatively modest but notable improvement over the GGA functional alone. The GGA$+U$ calculations effectively shift the results towards a better agreement with the experimental findings reported by Dewaele \textit{et al.}\cite{PhysRevLett.97.215504}. Moreover, our GGA$+U$ results are consistent with the findings obtained from LDA calculations augmented by dynamical mean field theory (DMFT) calculations, as reported by Pourvoskii \textit{et al.}\cite{pourovskii2019electronic}. However, our findings suggest that employing a GGA functional alone is sufficient to reasonably capture the ground state properties~\cite{cococcioni2005linear} of the ferromagnetic BCC phase, particularly when considering the significant static exchange splitting present in the system.

As pressure increases, the antiferromagnetic HCP phase becomes the stable phase~\cite{steinle2004magnetism,cohen2004non,lizarraga2008noncollinear,PhysRevB.67.180405}, starting at approximately 13 GPa (refer to Fig. \ref{enthalpy_fig} in Appendix~\ref{app} for the enthalpy curve). In this phase, electronic correlations play a more significant role as the static exchange splitting reduces bonding. Previous studies have considered these dynamical many-body effects using dynamical mean field theory (DMFT) to account for the impact of the +$U$ correction on the EOS~\cite{PhysRevB.90.155120}.

In the HCP phase, we observe a similar trend to the BCC phase. At lower pressures, HCP iron exhibits strong correlation effects. While our results obtained using GGA agree well with available DFT and quantum Monte Carlo (QMC) data in the literature~\cite{PhysRevB.81.094105,zhuang2021thermodynamic,sola2009equation,steinle2004magnetism_jpcm}, they visibly deviate from experimental measurements. Consequently, our GGA$+U$ calculations provide more accurate results and are in agreement with the LDA+DMFT results of Pourvoskii \textit{et al.}\cite{pourovskii2019electronic} in the pressure range of 13-30 GPa. However, even with the inclusion of the Hubbard correction to improve the treatment of strong correlations, the volume change in the EOS still deviates from experimental data\cite{boehler2008melting,PhysRevLett.97.215504,sakai2014equation,fei2016thermal,ono2010high} by approximately 5\% at around 70 a.u.$^{3}$/atom. The GGA predictions yield a $c/a$ ratio of less than 1.60, which further decreases with increasing volume~\cite{PhysRevB.90.155120}, while DMFT predicts a nearly constant $c/a$ ratio of approximately 1.60 regardless of volume, consistent with experimental measurements~\cite{PhysRevLett.110.117206,PhysRevLett.97.215504}. In the pressure range of 40-50 GPa, the correlation effects are slightly better captured by LDA+DMFT compared to GGA$+U$. At higher pressures (>100 GPa), our GGA results align well with experimental data~\cite{boehler2008melting,PhysRevLett.97.215504,sakai2014equation,fei2016thermal,ono2010high} and previous theoretical efforts~\cite{sha2011first,steinle2004magnetism_jpcm,zhuang2021thermodynamic} as the importance of correlations diminishes.

Fig. \ref{eos}b) shows the spin-resolved DOS of HCP iron at high pressure computed using DFT and DFT+$U$. At relatively low pressure (17.7~GPa), DFT+$U$ results in substantial changes in the occupancy of the $3d$ electrons near the Fermi energy. The DOS in the HCP phase at ambient pressure computed using DFT+DMFT~\cite{hausoel2017local} is additionally shown for reference. There is a striking resemblance in the peak feature above the Fermi energy between the DFT+$U$ and DMFT results, even though the DMFT results are calculated at ambient pressure. 

Fig. \ref{eos}c) shows the phonon DOS obtained via finite differences force constants as implemented in phonopy~\cite{phonopy-phono3py-JPCM,phonopy-phono3py-JPSJ} at pressures of 71.5~GPa and 138~GPa. They agree with the experimental results obtained using nuclear resonant inelastic x-ray scattering (NRIXS) at pressures of 71~GPa and 136~GPa respectively~\cite{gleason2013sound}. The figure clearly demonstrates that the influence of the Hubbard correction on the phonon density of states (DOS) is insignificant within the considered pressure ranges, consistent with the available experimental data.

\subsection{Dynamical Conductivity}
In the following, we investigate the impact of electronic correlation effects on the dynamical conductivity of iron in both the BCC and HCP phases.

\subsubsection{BCC Iron} 
\begin{figure}[b]
\centering   
\includegraphics[width=\linewidth]{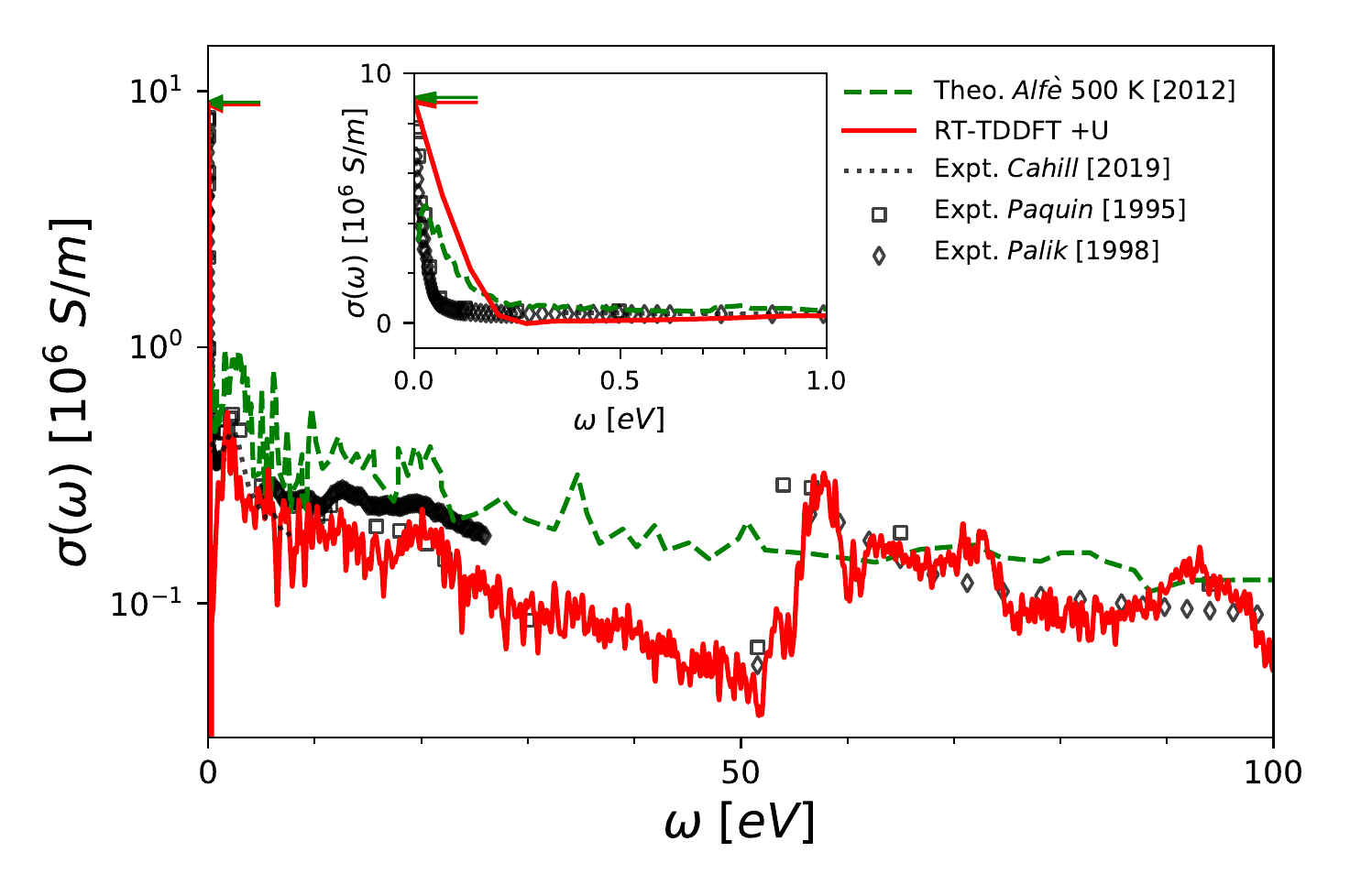} 
\caption{Dynamical electrical conductivity of BCC iron at ambient conditions computed using RT-TDDFT with Hubbard corrections. DC values for the RT-TDDFT (red) and KG (green) results are indicated by the arrows.
The inset plot shows the dynamical electrical conductivity at low energies. 
Experimental data stems from Refs.~\onlinecite{paquin1995properties,cahill2019optical,palik1998handbook}. Theoretical data at T=500~K evaluated using the Kubo-Greenwood formula stems from Ref.~\onlinecite{alfe2012lattice}. } 
\label{rt_tddft_ambient}   
\end{figure}

The frequency-dependent conductivity (Eq.~\ref{eq.ohmslaw.ft})  obtained using RT-TDDFT (red solid) including Hubbard corrections for BCC iron under ambient conditions is shown in Fig.~\ref{rt_tddft_ambient}. Due to the symmetry of the BCC lattice ($a = b = c$), it is sufficient to compute a single diagonal component of the conductivity tensor ($\sigma_{ab}$). In our case, we compute the $\sigma_{zz}$ component and compare our results with the KG results (green dashed) by Alf\`{e} \textit{et al.}~\cite{alfe2012lattice}, at a slightly elevated temperature (T=500~K), and experimental results by Paquain~\cite{paquin1995properties}, Palik~\cite{palik1998handbook} and Cahill~\cite{cahill2019optical} at ambient temperature. Overall, the RT-TDDFT results are in remarkable agreement with the experimental results by Paquain~\cite{paquin1995properties} and Palik~\cite{palik1998handbook} throughout the considered frequency range, and with the experimental results by Cahill~\cite{cahill2019optical} at lower frequencies. In contrast, the KG method, widely regarded as the standard approach for computing electrical conductivities, falls short of providing an accurate description across the entire frequency domain. In not displaying the drop in conductivity up to $50$~eV, the KG method averages over the feature at $\sim$55~eV which stems from \textit{3p} to conduction band transitions.

The DC conductivities obtained from the KG method (9.03~$\si{\mega \siemens \metre^{-1}}$), extracted through a Drude-like fit using different \emph{k}-point sampling and cell sizes~\cite{alfe2012lattice}, exhibit values close to those reported in experimental studies ($\sim$10~$\si{\mega \siemens \metre^{-1}}$)~\cite{gomi2013high,seagle2013electrical}. Similarly, the DC conductivities derived from RT-TDDFT+U yield a value of 8.83~$\si{\mega \siemens \metre^{-1}}$, further confirming their consistency with experimental observations.

It is common to verify simulation data in terms of the \textit{f}-sum rule~\cite{mahan2013many} 
\begin{equation}
\frac{1}{\pi}  \int_{0}^{\infty} d\omega\ \sigma(\omega)=  n_{e},
\end{equation}
where $n_{e}$ is the number of electrons in the simulation cell. Evaluated in a frequency range up to $\sim$5~Ha (i.e., $\sim$136~eV) and normalized to the \textit{f}-sum rule of the experimental data by Paquin~\cite{paquin1995properties}, this yields ratio of 0.95 for RT-TDDFT compared to 0.97 using the KG formula obtained by Alf\`{e} \textit{et al.}~\cite{alfe2012lattice} evaluated up to $\sim$5.5~Ha (i.e., $\sim$150~eV directly from the \textit{f}-sum rule) providing a measure of the quality of our calculations.  

\subsubsection{HCP Iron} 

\begin{figure}[t] 
\centering   
\includegraphics[width=\linewidth]{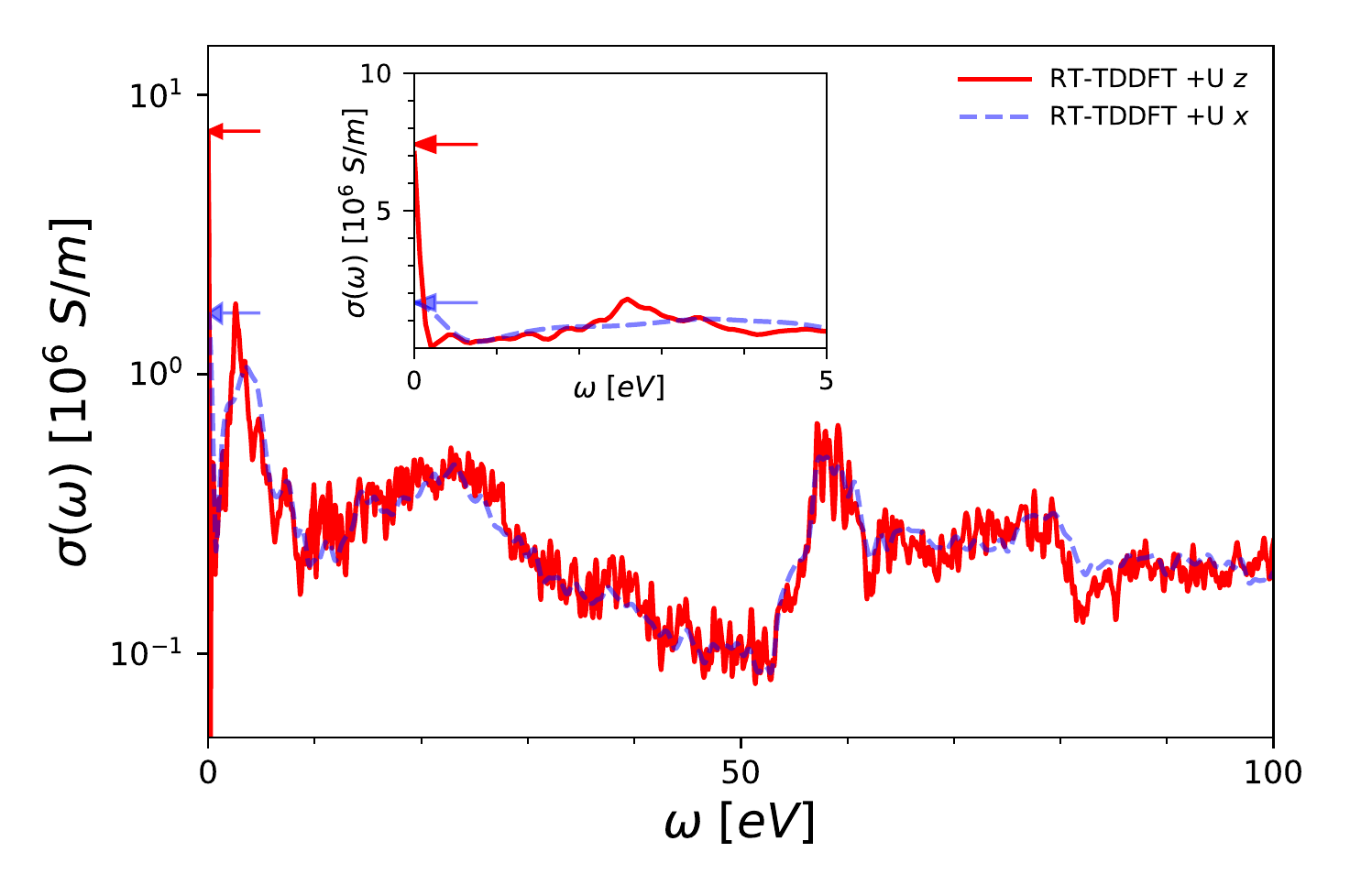} 
\caption{Dynamical electrical conductivity of HCP iron at 23~GPa computed using RT-TDDFT with Hubbard $U$ corrections along the $z$ (red) and $x$ (blue) directions. The inset plot shows the dynamical electrical conductivity at low energies. Anisotropy in the DC conductivity is inferred from the arrows.} \label{rt_tddft_ambient_hcp} 
\end{figure}

In the HCP phase, we evaluate the conductivity tensor similar to the BCC phase discussed earlier but along various directions and analyze the corresponding tensor components. 
This is relevant because the asymmetrical HCP lattice ($a = b \neq c$), as opposed to the BCC lattice, requires the computation of additional tensor components and could be of great aid for characterizing materials exhibiting the Hall effect~\cite{hall1879new,arxiv.2204.01538}. Fig.~\ref{rt_tddft_ambient_hcp} shows the frequency-dependent conductivity of HCP iron at 23~GPa obtained using RT-TDDFT with Hubbard corrections along the $z$ ($\sigma_{zz}$ - red solid) and $x$ ($\sigma_{xx}$ - blue dashed) directions. Notably, no experimental or theoretical data exist for direct comparison under these conditions, making our study the first benchmark for future experiments in this domain. 

The DC conductivities are represented by arrows, with the $z$ and $x$ directions yielding DC values ranging from 7.42~$\si{\mega \siemens \metre^{-1}}$ to 1.65~$\si{\mega \siemens \metre^{-1}}$, respectively. These values align with experimental resistivity (inverse of conductivity) measurements~\cite{seagle2013electrical,gomi2013high}. The significant anisotropy in the DC conductivity holds particular relevance for experiments conducted with diamond anvil cells (DAC), as crystallographic orientations play a vital role in determining the observed behavior.

\subsection{DC Conductivity with Pressure} 

\begin{figure}[t] 
\centering       
\includegraphics[width=\linewidth]{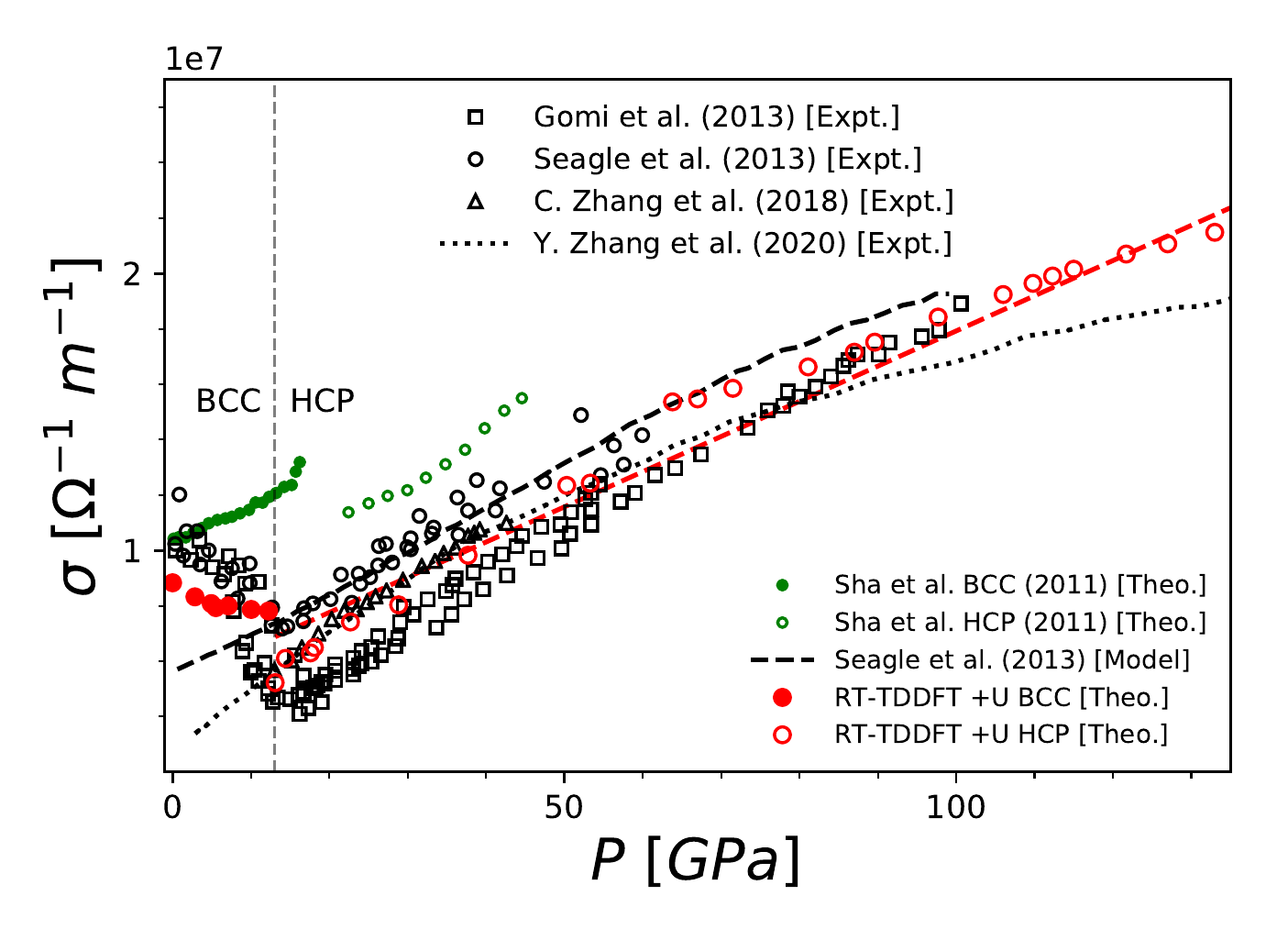}
\caption{Pressure versus DC conductivity at ambient temperature (T=300~K). The results of this work are indicated by solid and empty red circles for BCC and HCP phases respectively. The drop in the electrical conductivity around $\sim$13~GPa indicates the BCC$\rightarrow$HCP phase transition. The red dashed line is the linear fit to our HCP data. The black dashed line is a pressure and temperature-based model by Seagle \textit{et al.}~\cite{seagle2013electrical}. The black dotted line is the fitted results to the experimental measurements by Y. Zhang \textit{et al.}~\cite{zhang2020}.  Theoretical data stems from Ref.~\onlinecite{sha2011first}. Experimental data stems from Refs.~\onlinecite{gomi2013high,seagle2013electrical,10.1029/2017JB015260}.}
\label{sigma_300K}
\end{figure}

Now we turn to the DC conductivity evaluated using RT-TDDFT at high pressures. Fig.~\ref{sigma_300K} shows the DC conductivity as a function of pressure in the BCC and HCP phases at ambient temperature (T=300~K). The results in the BCC phase are indicated by solid red circles and those in the HCP phase by empty red circles. In the BCC phase, our predicted DC conductivity is in the range reported by several experiments~\cite{yousuf1986high,gomi2013high,paquin1995properties,palik1998handbook}. With increasing pressure in the BCC phase, the DC conductivity drops in agreement with the experiments~\cite{gomi2013high,seagle2013electrical}. There is a further drop in the conductivity observed near 13~GPa which marks the transition between the BCC and the HCP phase. Note that a similar drop in the conductivity during the phase transition has been observed for several other materials~\cite{balchan1961high,garg2004electrical}. In contrast to the experimental results~\cite{gomi2013high,seagle2013electrical} and our results, \textit{ab initio} calculations by Sha \textit{et al.}~\cite{sha2011first} report an increase in the conductivity for the BCC phase with pressure although the drop in the conductivity during the phase transition is captured. The discrepancies in the theoretical results can be attributed to the use of density functional perturbation theory (DFPT) by Sha \textit{et al.} to evaluate the electrical resistivity, whereas our work employs the more effective approach of RT-TDDFT. By considering electronic correlations more accurately, as demonstrated earlier in the analysis of dynamical electrical conductivity, RT-TDDFT proves to be a superior method for capturing the relevant physics.

Our results in the HCP phase exhibit good agreement with experimental data from Seagle and Gomi \textit{et al.}\cite{seagle2013electrical,gomi2013high}, as well as with the findings of C. Zhang \textit{et al.}\cite{10.1029/2017JB015260} up to 50~GPa. Additionally, our results align well with the measurements on HCP iron by Y. Zhang \textit{et al.}\cite{zhang2020}, as indicated by the black dotted line, particularly for pressures below 50~GPa. Notably, our results differ significantly from the calculations by Sha \textit{et al.}\cite{sha2011first}, as their estimations overestimate the conductivity in the HCP phase. The black dashed line represents the model-based pressure and temperature dependence of the resistivity, derived by Seagle \textit{et al.}\cite{seagle2013electrical}, which agrees well with our findings for pressures above 50~GPa.

\subsection{Thermal Conductivity in HCP Iron with Pressure}  

\begin{figure}[t]    
\centering       
\includegraphics[width=\linewidth]{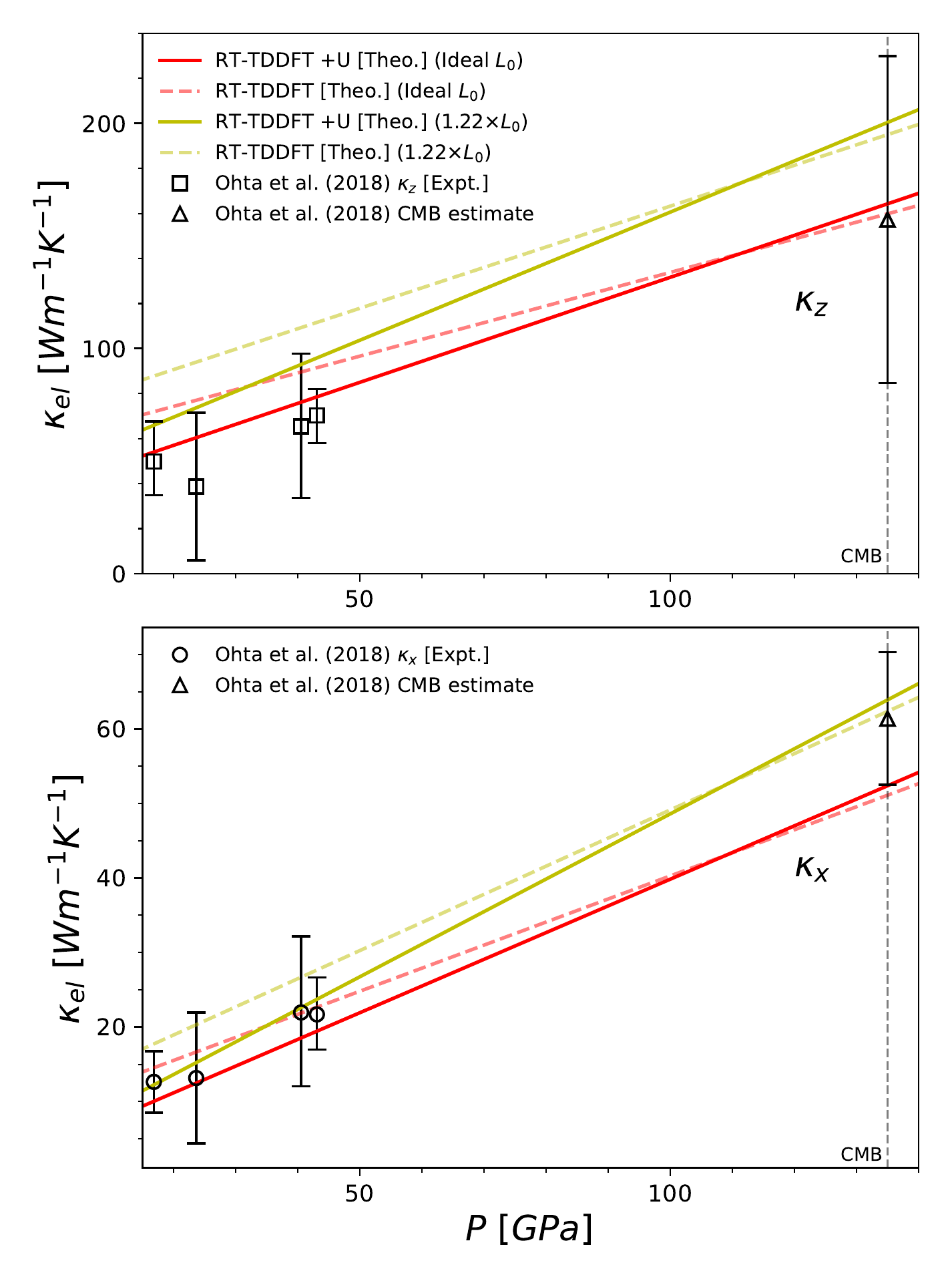}   
\caption{Electronic components of the thermal conductivity of HCP iron along different directions (top panel - $z$ direction, bottom panel - $x$ direction) vs pressure. The inclusion of Hubbard corrections ($+U$) is additionally shown. The dashed and the solid lines are the linear fit to RT-TDDFT and RT-TDDFT+$U$ results respectively. The red and the yellow curves shown are computed using the ideal Lorenz number ($L_{0}$) and  1.22$\times L_{0}$ respectively. Experimental anisotropy data (black hollow squares and black hollow circles) at ambient temperature (T=300~K) stems from Ref.~\onlinecite{ohta2018experimental}. Estimated data (black hollow triangles) at the core-mantle boundary (CMB) conditions from Ref.~\onlinecite{ohta2018experimental}.}   
\label{anisotropy}  
\end{figure}   

Finally, we analyze the electronic component of the thermal conductivity of HCP iron along the $z$ (top panel) and $x$ (bottom panel) directions, as depicted in Fig.~\ref{anisotropy}. The thermal conductivity is determined from the electrical conductivity using the Wiedemann-Franz law~\cite{franz1853ueber}. Given the pressure and temperature conditions in this study, we expect the Wiedemann-Franz law to be valid, as both energy and charge transport is predominantly governed by the same free carriers, with inelastic scattering effects becoming less significant. For our calculations, we employ the ideal Lorenz number ($L_{0}$=2.44$\times 10^{-8}$~W$\Omega$K$^{-2}$), as well as an experimentally determined Lorenz number (1.22$\times L_{0}$) established by Secco \textit{et al.}\cite{secco2017thermal} under ambient conditions, for comparison. It is worth noting that the conventional Lorenz number used under high-pressure and high-temperature conditions exhibits substantial variations across previous studies~\cite{pozzo2012thermal,de2012electrical,pozzo2014thermal,ohta2016experimental,konopkova2016direct,PhysRevLett.121.096601}.

Our findings exhibit excellent agreement with the experimental data obtained from laser-heated diamond anvil cell (DAC) measurements conducted by Ohta \textit{et al.}\cite{ohta2018experimental} (depicted as black hollow squares and black hollow circles for $\kappa_{z}$ and $\kappa_{x}$, respectively) at pressures below 50~GPa. Notably, the Hubbard corrections play a crucial role, particularly at lower pressures where electronic correlations are more pronounced. Neglecting the Hubbard corrections leads to a deviation of up to 40$\%$ along the $z$ direction around 13~GPa. We extend our calculations to the pressure regime encountered at the core-mantle boundary (CMB) in Earth's interior (P$\sim$135~GPa). Here, we additionally present the estimated thermal conductivity (depicted as black hollow triangles) by Ohta~\cite{ohta2018experimental}. At CMB conditions, we report a $\kappa_{z}/ \kappa_{x}$ ratio of 3.03, which is comparable to the value of $2.56 \pm 0.48$ estimated by Ohta \textit{et al.}\cite{ohta2018experimental}. The thermal conductivity of pure iron at CMB conditions is estimated by Seagle \textit{et al.}\cite{seagle2013electrical} to be 145~Wm$^{-1}$K$^{-1}$. Our predictions fall within this range, with a value of $158.4$~Wm$^{-1}$K$^{-1}$ ($197.4$~Wm$^{-1}$K$^{-1}$ when considering 1.22$\times L_{0}$), aligning with previous theoretical efforts\cite{de2012electrical,pozzo2012thermal}. Taking the trace average of the conductivity tensor $ \sum_{i=(x,y,z)} \sigma_{ii} / 3$ into account, the resulting thermal conductivity near CMB conditions (P$\sim$135GPa) is approximately 87~Wm$^{-1}$K$^{-1}$ (approximately 109~Wm$^{-1}$K$^{-1}$ when considering 1.22$\times L_{0}$).
This analysis emphasizes the impact of electronic correlations on thermal conductivity at pressures up to 50 GPa, underscoring the importance of employing the RT-TDDFT methodology with Hubbard corrections for accurate simulation and modeling.  

In recent studies, the total thermal conductivity, accounting, in addition, for the coupling between phonons and magnons, has been split into various contributions, as demonstrated by Nikolov \textit{et al.}\cite{nikolov2022dissociating}. Their work highlights the significant contribution of the electronic component (approximately 80-90$\%$) at ambient temperature\cite{backlund1961experimental,wu2018magnon}. For the purpose of this study, we have thus focused solely on the electronic contribution to thermal conductivity. Experimental measurements of ferromagnetic BCC iron at ambient conditions show a total thermal conductivity of 77.3~Wm$^{-1}$K$^{-1}$, with the electronic contribution estimated at 69.4~Wm$^{-1}$K$^{-1}$ according to the Wiedemann-Franz law~\cite{fulkerson1966comparison}. In comparison, our calculations yield a thermal conductivity of 64.6~Wm$^{-1}$K$^{-1}$ (78.6~Wm$^{-1}$K$^{-1}$ using 1.22$\times L_{0}$) according to the Wiedemann-Franz law.

\section{Conclusions} 

In this study, we comprehensively investigated the influence of electronic correlations on various properties of solid iron at high pressures. These properties include ground state characteristics such as the EOS, the electronic DOS, and the phonon DOS, as well as electronic transport properties including electrical and thermal conductivity. To perform our analysis, we employed RT-TDDFT with Hubbard corrections and compared the results with those obtained using the conventional KG method for transport properties. 

Our findings clearly demonstrate that electronic correlation effects have a significant impact on the investigated properties up to pressures of 50 GPa~\cite{PhysRevB.90.155120}. Moreover, our work highlights the advantages of utilizing RT-TDDFT (including Hubbard corrections) as a more reliable method for computing electronic transport properties compared to the current state-of-the-art approach, such as the KG method.

We envision promising applications of this methodology in studying structural phase transitions and spin response~\cite{kaa2022structural} in materials under extreme conditions, made possible by recent advancements in free-electron lasers~\cite{zastrau2021high,cerantola2021new}. The frequency-dependent electrical conductivity, along with the analysis of anisotropy in HCP iron, can serve as a valuable benchmark for future experiments, particularly in the context of advanced diagnostics.

Furthermore, the role of magnetism in iron-nickel alloys under Earth-core conditions has been shown to be significant, highlighting the importance of electronic correlations, which were previously assumed to be negligible~\cite{vekilova2015electronic,hausoel2017local}. In the future, incorporating electronic correlations into the \textit{ab initio} analysis of electronic transport properties in the liquid phase of iron and iron alloys relevant to Earth-core conditions would be an important task to pursue.

\begin{acknowledgments}
This work was partially supported by the Center for Advanced Systems Understanding (CASUS) which is financed by Germany’s Federal Ministry of Education and Research (BMBF) and by the Saxon state government out of the State budget approved by the Saxon State Parliament. Computations were performed on a Bull Cluster at the Center for Information Services and High-Performance Computing (ZIH) at Technische Universit\"at Dresden and on the cluster Hemera at Helmholtz-Zentrum Dresden-Rossendorf (HZDR). 
Sandia National Laboratories is a multimission laboratory managed and operated by National Technology \& Engineering Solutions of Sandia, LLC, a wholly-owned subsidiary of Honeywell International Inc., for the U.S. Department of Energy’s National Nuclear Security Administration under contract DE-NA0003525. This paper describes objective technical results and analysis. Any subjective views or opinions that might be expressed in the paper do not necessarily represent the views of the U.S. Department of Energy or the United States Government. 
\end{acknowledgments}

\section*{Data Availability}
The data that support the findings of this study are available from the corresponding author upon reasonable request.

\appendix

\section{Appendix}\label{app}

A comparison between the calculated lattice constant, magnetic moment, and bulk modulus within several approximate DFT schemes, \textit{ab initio} methods, and experimental results are shown in Table. \ref{Fe_comp_table}. 

\begin{table} 
\caption{Comparison between the calculated lattice constant, magnetic moment, and  bulk modulus within several approximate DFT schemes, experimental and theoretical results for BCC iron.}
\begin{ruledtabular}
\begin{tabular}{lccc} 
Method & $a_{0}$~(a.u.) & $\mu_{0}$~($\mu_{b}$) & $B_{0}$~(GPa)  \\
\hline  
Spin-pol. LDA & 5.21 & 2.02 & 259 \\
Spin-pol. PBEsol & 5.28 & 2.09 & 213 \\
Spin-pol. PBE & 5.37 & 2.19 & 169 \\
Spin-pol. PBE+$U$  & 5.39 & - & 166  \\
Spin-pol. LDA$^{a}$  & 5.20 & 1.95 & - \\
Spin-pol. LDA+$U^{a}$  & 5.35 & 2.73 & - \\
Spin-pol. PBEsol$^{a}$  & 5.35 & 2.12 & - \\
Spin-pol. PBEsol+$U^{a}$  & 5.35 & 2.71 & - \\
Spin-pol. PBE$^{a}$  & 5.20 & 2.19 & - \\
Spin-pol. PBE+$U^{a}$  & 5.37 & 2.09 & - \\
Spin-pol. LDA$^{e}$ & 5.22 & 2.33 & 210 \\
LDA+DMFT$^{f}$ & 5.39 & - & 172 \\
Expt. & 5.42$^{b}$ & 2.22$^{c}$ & 173$^{d}$ \\ 
\end{tabular}   
\end{ruledtabular}
\footnotetext[1]{Ref.~\onlinecite{tavadze2021exploring}.} \footnotetext[2]{Ref.~\onlinecite{chiarotti19951}.}
\footnotetext[3]{Ref.~\onlinecite{kikuchi1990structure}.}
\footnotetext[4]{Ref.~\onlinecite{adams2006elastic}.}
\footnotetext[5]{Ref.~\onlinecite{cococcioni2005linear}.}
\footnotetext[6]{Ref.~\onlinecite{PhysRevB.90.155120}.}
\label{Fe_comp_table}  
\end{table}   

\begin{figure}
\centering       
\includegraphics[width=\linewidth]{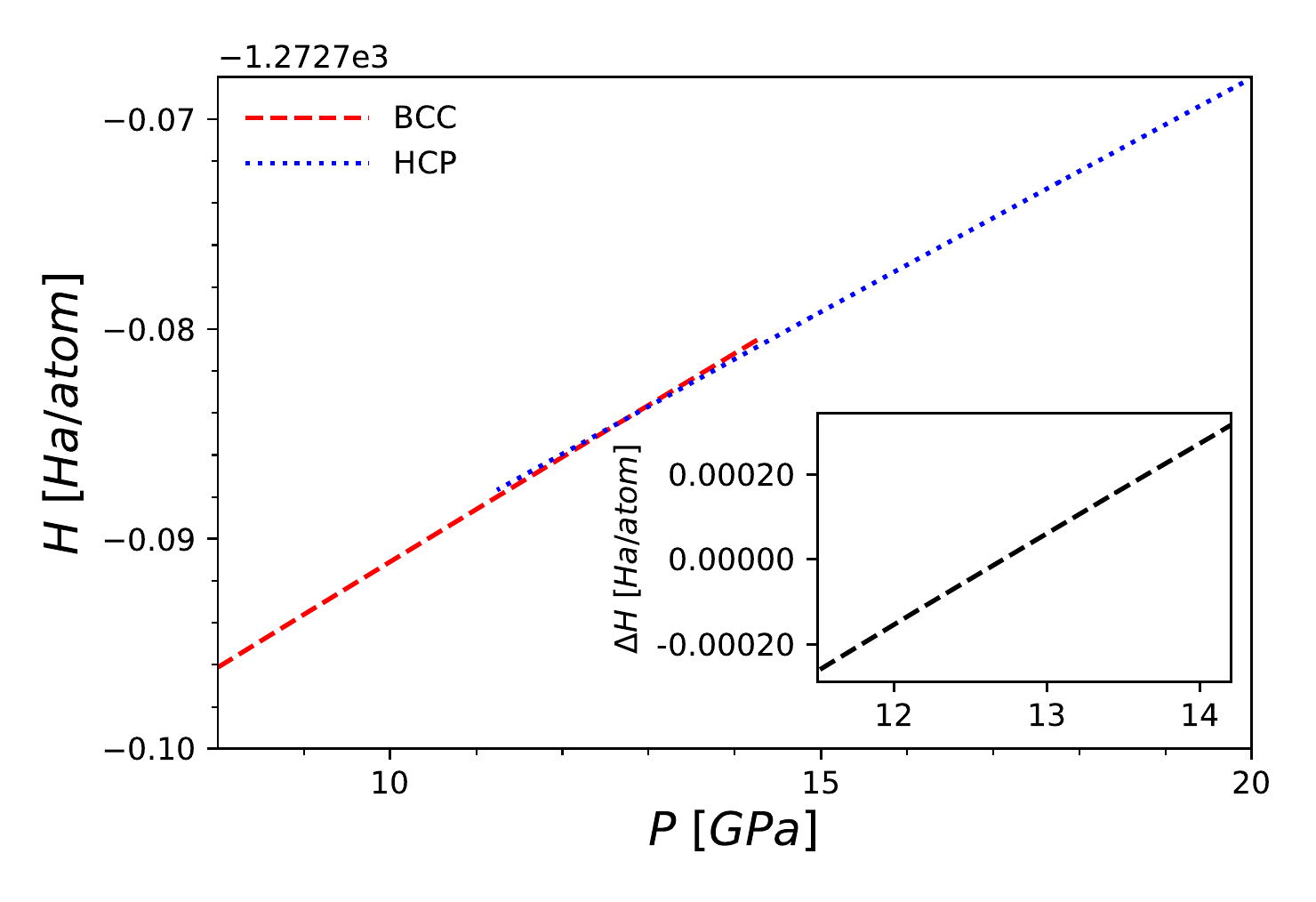}
\caption{Enthalpy versus pressure for BCC and HCP phases of iron. The inset plot shows the relative difference in enthalpy ($\Delta H_{BCC} - \Delta H_{HCP}$) between the phases versus pressure.}  
\label{enthalpy_fig}
\end{figure}

Fig. \ref{enthalpy_fig} shows the variation of enthalpy with pressure for the BCC and HCP phases of iron. 
The inset plot indicates the relative difference in enthalpy between the phases with the phase transition occurring at $\sim$12.9~GPa in agreement with the range observed in experiments~\cite{bancroft1956polymorphism,takahashi1964high}.

\clearpage

\nocite{*}
\bibliography{aipsamp}

\end{document}